\documentclass[twocolumn,letterpaper,accepted=2025-04-15]{quantumarticle}
\pdfoutput=1
\usepackage{graphicx}
\usepackage{dcolumn}
\usepackage{bm}
\usepackage{braket}
\usepackage{amsmath}
\usepackage{xcolor}
\usepackage{algorithm}
\usepackage{algorithmic}
\usepackage{qcircuit}
\usepackage{hyperref}

\begin{document}

\title{Efficient separate quantification of state preparation errors and measurement errors on quantum computers and their mitigation}

\author{Hongye Yu}
\affiliation{C. N. Yang Institute for Theoretical Physics,
State University of New York at
Stony Brook, Stony Brook, NY 11794-3840, USA}
\affiliation{Department of Physics and Astronomy, State University of New York at
Stony Brook, Stony Brook, NY 11794-3800, USA}
\author{Tzu-Chieh Wei}
\affiliation{C. N. Yang Institute for Theoretical Physics,
State University of New York at
Stony Brook, Stony Brook, NY 11794-3840, USA}
\affiliation{Department of Physics and Astronomy, State University of New York at
Stony Brook, Stony Brook, NY 11794-3800, USA}
\begin{abstract}
Current noisy quantum computers have multiple types of errors, which can occur in the state preparation, measurement/readout, and gate operations, as well as intrinsic decoherence and relaxation. Partly motivated by the booming of intermediate-scale quantum processors, measurement and gate errors have recently been extensively studied, and several methods of mitigating them have been proposed and formulated in software packages (e.g., IBM Qiskit). Despite this, the state preparation error and the procedure to quantify it have not yet been standardized, as state preparation and measurement errors are usually considered to be not directly separable.  Inspired by a recent work of Laflamme, Lin, and Mor [Phys. Rev. A {\bf 106}, 012439 (2022)], we propose a simple and resource-efficient approach to quantify separately the state preparation and readout error rates. With these two errors separately quantified, we also propose methods to mitigate them separately, especially mitigating state preparation errors with linear (with the number of qubits) complexity. As a result of the separate mitigation, we show that the fidelity of the outcome can be improved by an order of magnitude compared to the standard measurement error mitigation scheme. We also show that the quantification and mitigation schemes are resilient against gate noises and can be immediately applied to current noisy quantum computers. To demonstrate this, we present results from cloud experiments on IBM's superconducting quantum computers. The results indicate that the state preparation error rate is also an important metric for qubit metrology that can be efficiently obtained.
\end{abstract}

\maketitle

\section{Introduction}
There has been dramatic growth in multi-qubit quantum processors; some possess qubits well over one hundred. Nevertheless, they are characterized as noisy-intermediate quantum (NISQ) devices~\cite{preskill2018quantum} as it is currently not yet feasible to implement full-scale quantum error correction on these devices to reduce logical error rates and prolong coherent computation. Hence, a short-term goal is to develop error mitigation schemes to enhance the performance of NISQ devices. Errors and noise arise in different stages of quantum computation: state preparation, gate operation, and readout/measurement. In the past few years, substantial efforts have been spent on gate error mitigation~\cite{temme2017error,endo2018practical,dumitrescu2018cloud,kandala2019error,giurgica2020digital,kim2023scalable,van2023probabilistic}, as gate operation is the bulk of quantum computation. 

 To mitigate readout/measurement errors, several methods have been proposed~\cite{chen2019detector,maciejewski2020mitigation,geller2021toward,bravyi2021mitigating,van2022model} to mitigate the final readout statistics, and have improved the estimation of physical observables from quantum computers.  State preparation is perhaps the most omitted among the three types of errors. At present, there does not seem to be an efficient and practical approach to even quantifying state preparation and readout errors separately. Usually, the state preparation error and measurement error are considered inseparable and are combined as the state preparation and measurement (SPAM) error. Indeed, if we only have one qubit, prepared at $\ket{0}$ but measured with outcome $\ket{1}$, we cannot tell whether the outcome error comes from the stage in the measurement or the initial state preparation. Moreover, current measurement mitigation schemes on superconducting qubits, such as in the Qiskit software~\cite{Qiskit}, are still mixed with the state preparation error. Although the rate of state-preparation error is typically lower than that of the readout error, for quantum computers with state-preparation error dominating, such mixed mitigation can have large systematic errors proportional to the error rate of the state preparation.

The goal of this work is two-fold. First, we propose a simple and efficient scheme that allows separate quantification of state preparation and measurement errors. This issue was recently discussed by Laflamme, Lin, and Mor~\cite{laflamme2022algorithmic} employing algorithmic cooling~\cite{baugh2005experimental,ryan2008spin,schulman2005physical}. Our approach extends theirs, but dramatically simplifies the procedure. Both schemes rely on coupling the original qubit to other qubits.  Ours requires only one additional qubit to characterize the errors in state preparation and measurement separately. Another approach to characterize gates and SPAM errors with very few assumptions is gate set tomography~\cite{blume2017demonstration}; however, it requires many more gate sequences and is not feasible for large-system experiments, whereas ours can be easily implemented on current noisy quantum computers even for large numbers of qubits. Secondly, with separate quantification, the measurement error, in principle, is free of the state preparation error, and the current standard mitigation scheme can be directly applied to mitigating the measurement error. Moreover, we propose a novel approach to mitigate the state-preparation error. The mitigation scheme is practical and requires an $O(N)$ overhead for first-order full-qubit mitigation. The scheme can also be done qubit-wisely, and thus the overhead can be greatly reduced if very few qubits (with relatively large state-preparation errors) need mitigation. The mitigation scheme can also be directly applied to mitigating expectation values of observables, not merely to the outcome probability distributions.
Even though we focus our attention and implementation on superconducting qubits, we envision that our method can be applied to other systems as well. 

We also remark that in this work, we assume that single-qubit gates, in particular, the X gate, are of high fidelity in the characterization circuits. The assumption is reasonable in practice. In current superconducting quantum computers~\cite{kim2023evidence}, the errors of single-qubit gates are around $10^{-4}$ while the SPAM error of qubits can be $10^{-2}$. We do not assume that the CNOT gate is noiseless, but we assume that existing error-mitigation techniques, such as zero-noise extrapolation (ZNE), can mitigate the CNOT gate error~\cite{temme2017error,endo2018practical,dumitrescu2018cloud,kandala2019error,giurgica2020digital,kim2023scalable} to sufficient accuracy if needed.

The remainder of the paper is organized as follows. In Sec.~\ref{sec:separatecharacterization}, we present a simple scheme using an ancilla to characterize the target qubit's state preparation and measurement errors separately. 
In Sec.~\ref{sec:completemitigation}, we discuss how the characterized state preparation and readout errors can 
be separately mitigated. 
In Sec.~\ref{sec:comparison}, we compare our proposed mitigation scheme with the standard readout mitigation (which does not characterize the state preparation error) and illustrate the difference in mitigated outcomes for simple circuits.
To showcase the practicalness of our separate characterization and mitigation schemes, we present results from experiments performed on cloud IBM quantum computers to demonstrate our theoretical proposal in Sec.~\ref{sec:separatecharacterization} and ~\ref{sec:comparison}. We make concluding remarks in Sec.~\ref{sec;conclusion}. 


\section{Separate characterization of state preparation and readout errors}
\label{sec:separatecharacterization}

A single-qubit readout can be characterized by a two-element POVM $\{M_0,M_1\}$, so that a one-qubit state $\rho$ will be measured to be `0' with  probability ${\rm Tr}(\rho M_0)$ and `1' with probability ${\rm Tr}(\rho M_1)$. The readout errors in practice are dominantly classical flips, i.e., the POVM elements are well approximated by the following  forms,
 \begin{equation}
 M_0=\begin{pmatrix}
     1-\delta_M^0 & 0\\
     0 & \delta_M^1
 \end{pmatrix}, \  \ M_1=\begin{pmatrix}
     \delta_M^0 & 0\\
     0 & 1-\delta_M^1
 \end{pmatrix},
 \end{equation}
where the parameter $\delta_M^0$ is the probability that the ideal `0' state will be read as `1', and $\delta_M^1$ is the probability that the ideal `1' will be read as `0'; see, e.g., Ref.~\cite{chen2019detector}, which also demonstrated this on superconducting qubits of IBM's devices. We note that the POVM for the general single-qubit readout can be more general than the above form, but it can be enforced to the above form by twirling with a random Pauli Z operator (with 50\% probability) before the measurement~\cite{chen2019detector,laflamme2022algorithmic}. 

In Ref.~\cite{laflamme2022algorithmic}, a measurement-based algorithmic cooling was developed to reduce the state-preparation error, where  the error ruins the preparation of a perfect $|0\rangle$, i.e.,
\begin{equation}
 \label{eq:SP}
\rho_0^{[\rm err]} \equiv (1-\delta_{\rm SP}) |0\rangle\langle0|+ \delta_{\rm SP}|1\rangle\langle 1|,
\end{equation}
and the cooling was achieved by coupling to ancillary qubits via CNOTs. There, symmetric readout errors were considered, i.e., $\delta_M^0=\delta_M^1=:\delta_{\rm M}$ and a procedure employing multiple ancillary qubits was proposed to separately characterize a target qubit's $\delta_{\rm SP}$ and $\delta_{\rm M}$ (in the limit of many ancillas), which comprises the total SPAM error $\delta_{\rm SPAM}=\delta_{\rm SP}+\delta_{\rm M}-2\delta_{\rm SP}\delta_{\rm M}$, giving the measurable probability of reading out `1'. The above result can be easily derived from $\delta_{\rm SPAM}={\rm Tr}(M_1 \rho_0^{[\rm err]})$ with symmetric error in $M_i$'s.  We give further detailed discussions related to the algorithmic cooling in Appendix.~\ref{app:LLM}. Naively, one might have expected that the total contribution of the SPAM error is simply the sum of two errors. However, it is important to note that the total SPAM error does contain the crossed term $-2\delta_{\rm SP}\delta_{\rm M}$ that reduces the sum of the two. Thus, it is clear that \textit{using $\delta_{\rm SPAM}$ to perform readout mitigation will contain the error from state preparation}, which, however, is the current standard approach.

The above discussion, therefore, motivates us to propose a simplified procedure in this section, using one ancilla to extract the state-preparation error $\delta_{\rm SP}$ from readout errors $\delta_M^0$ and $\delta_M^1$ separately. First, we note that a coherent state preparation error, e.g., $\sqrt{1-\delta_{\rm SP}}|0\rangle+\sqrt{\delta_{\rm SP}}e^{i\phi}|1\rangle$, can be turned into the above incoherent state preparation error~(\ref{eq:SP}) by twirling with a Pauli Z operator randomly with 50\% probability~\cite{laflamme2022algorithmic}. We note that the coherent error can be probed using quantum state tomography once the readout errors have been characterized and mitigated; see Sec.~\ref{sec:tomography} below. We will not assume the symmetric readout errors but actively employ the twirling to ensure the incoherent form of state-preparation error. The parameters $\delta_M^0$ and $\delta_M^1$ are reported on the device properties of all IBM quantum computers, except that the reported values are mixed with the state preparation error. Can the true readout errors be obtained experimentally and separately from the state preparation error? Indeed, Ref.~\cite{laflamme2022algorithmic} describes an approach by algorithmic cooling to reduce the state preparation error, and in the limit that it becomes infinitesimally small, only the readout errors remain. However, the reduction to zero state preparation error is not practical as the required CNOT gates (which were assumed to be error-free in Ref.~\cite{laflamme2022algorithmic}) will introduce additional errors in practice. Our procedure below solves all these problems and can practically improve the device property characterization,  separating the state preparation and readout errors.

\subsection{Access $\delta_{sp}$ via noiseless CNOT gates}\label{sec:dsp}

\begin{figure}[h]
    \centering
    
    \[\Qcircuit @C=.7em @R=.7em @! {
\lstick{\ket{0}_\text{a}}& \targ     & \meter  & \rstick{\tilde{\delta}_{\rm SPAM}^{a,0}} \qw\\
\lstick{\ket{0}_\text{t}}& \ctrl{-1}\qw  & \meter  & \rstick{\delta_{\rm SPAM}^{t,0}} \qw
}\]

\[\Qcircuit @C=.7em @R=.7em @! {
\lstick{\ket{0}_\text{a}}&\qw& \targ     & \meter  & \rstick{\tilde{\delta}_{\rm SPAM}^{a,1}} \qw\\
\lstick{\ket{0}_\text{t}}&\gate{X}& \ctrl{-1}\qw  & \meter  & \rstick{\delta_{\rm SPAM}^{t,1}} \qw
}\]
    \caption{Circuits for measuring $\delta_{\rm SP}$ with initial state prepared to be: (top) $\ket{00}$ and (bottom) $\ket{01}$. Here we assume the X gate is noiseless and can be considered as a part of state preparation.}
    \label{fig:circ_dsp}
\end{figure}

Here we present an efficient way to estimate the state-preparation error $\delta_{\rm SP}^t$ for the target qubit $q_t$, which only requires one additional qubit $q_a$, as shown in the Fig.~\ref{fig:circ_dsp}. We remark that such circuits are similar to a repetition code, which has been applied to suppressing both state-preparation errors~\cite{laflamme2022algorithmic} and readout errors~\cite{gunther2021improving,hicks2022active}. In this work, the circuits are only applied to the quantification process, not the active suppression process.

In these circuits, the SPAM error of both qubits can be easily measured. To measure the probability $\delta_{\rm SPAM}^{0}$ that a qubit is prepared with $\ket{0}$ but measured in $\ket{1}$, one only needs to measure the initialized qubit directly; to measure the probability $\delta_{\rm SPAM}^{1}$ that the qubit is prepared with $\ket{1}$ but measured in $\ket{0}$, one needs to first initialize the qubit in $\ket{0}$,  apply an X gate, and then measure it. As emphasized above, the measured error $\delta_{\rm SPAM}$ is a combination of both state-preparation error $\delta_{\rm SP}$ and measurement errors $\delta_{M}$'s, and, without assuming symmetric readout errors, we have $\delta_{\rm SPAM}^{0}= \delta_M^{0}(1-\delta_{\rm SP})+(1-\delta_M^{1}) \delta_{\rm SP}$. Combined with similar derivations for $\delta_{\rm SPAM}^{1}$, we have
\begin{equation}\label{eq:SPAM=SP+M}
\begin{aligned}
    \delta_{\rm SPAM}^{0}= (1-\delta_M^{0}-\delta_M^{1}) \delta_{\rm SP} + \delta_M^{0},\\
    \delta_{\rm SPAM}^{1}= (1-\delta_M^{0}-\delta_M^{1}) \delta_{\rm SP} + \delta_M^{1},
\end{aligned}
\end{equation}
or equivalently,
\begin{equation}
    A_{\rm SPAM}=A_M A_{\rm SP},
\end{equation}
where $A$'s are the corresponding error assignment matrices (see Appendix~\ref{app:vecErr} for explicit definitions and derivations)
\begin{equation}
    A\equiv\begin{pmatrix}
        1-\delta^0 & \delta^1\\
        \delta^0 & 1-\delta^1
    \end{pmatrix}.
\end{equation}

 After adding a noiseless CNOT gate (Fig.~\ref{fig:circ_dsp}), the SPAM error of $q_t$ does not change, whereas the ancilla qubit readout error $\tilde{\delta}_{\rm SPAM}^{a,0/1}$ becomes 
\begin{equation}\label{eq:spamT}
\begin{aligned}
    \tilde{\delta}_{\rm SPAM}^{a,0}= (1-\delta_{\rm SPAM}^{a,0}-\delta_{\rm SPAM}^{a,1}) \delta_{\rm SP}^t + \delta_{\rm SPAM}^{a,0},\\
    \tilde{\delta}_{\rm SPAM}^{a,1}= (1-\delta_{\rm SPAM}^{a,0}-\delta_{\rm SPAM}^{a,1}) \delta_{\rm SP}^t + \delta_{\rm SPAM}^{a,1},
\end{aligned}
\end{equation}
where the superscript 0/1 indicates the ideal output and notation $\delta^{0/1}$'s indicate their error probabilities. The left-hand sides (emphasized by \~{}) are the results with a CNOT present, and the right-handed sides are results obtained in single-qubit circuits. 

These relations can be derived as follows. From the top panel of Fig.~\ref{fig:circ_dsp}, $\tilde{\delta}_{\rm SPAM}^{a,0}$ is the probability of measuring `1' on qubit $a$, which has two contributions due to noisy preparation of qubit $t$: (a) With probably $(1-\delta_{\rm SP}^t)$ the qubit $t$ is prepared in $|0\rangle$ and the CNOT has no effect; thus for qubit $a$, we are interested in the matrix element $(A_{\rm SPAM}^a)_{10}$ (i.e., preparing `0' and reading out `1'). (b) With probably $\delta_{\rm SP}^t$ the qubit $t$ is prepared in $|1\rangle$ and the CNOT has the effect of flipping qubit $a$; thus for qubit $a$, we are interested in the matrix element $(A_{\rm SPAM}^a)_{11}$.
This gives the first equation in Eq.~(\ref{eq:spamT}) via: $\tilde{\delta}_{\rm SPAM}^{a,0}=(1-\delta_{\rm SP}^t)(A_{\rm SPAM}^a)_{10}+\delta_{\rm SP}^t(A_{\rm SPAM}^a)_{11}$. Likewise, we have $\tilde{\delta}_{\rm SPAM}^{a,1}=(1-\delta_{\rm SP}^t)(A_{\rm SPAM}^a)_{01}+\delta_{\rm SP}^t(A_{\rm SPAM}^a)_{00}$. Actually, these arguments can be captured in the single matrix equation:
\begin{equation}
    \tilde{A}_{\rm SPAM}^a=A_{\rm SPAM}^a A_{\rm SP}^t;
\end{equation}
see also Appendix~\ref{app:vecErr} for slightly formal notation that streamlines the derivation.

By using either of the two equations in \eqref{eq:spamT} and the measured values of $\delta_{\rm SPAM}^{a, 0/1}$, one can solve the $\delta_{\rm SP}^t$.  Combined with the measured SPAM errors from $q_t$, one can further obtain the value of measurement errors $\delta_M^{0/1}$ for qubit $q_t$. 

We note that solving $\delta_M$'s and $\delta_{\rm SP}$ here involves only linear equations. One can also extend the algorithmic cooling of Laflamme, Lin, and Mor in the two-qubit setting to solve for (the symmetric) $\delta_M$ and $\delta_{\rm SP}$, but this will involve non-linear equations; see Appendix~\ref{app:LLM}. In the next section, we expand our analysis to noisy CNOT gates.

\subsection{Characterization of $\delta_{\rm SP}$ with noisy CNOT gates}
\label{sec:dspnoise}
As the CNOT gate noise is inevitable in current noisy quantum devices, directly applying the characterization circuits (Fig.~\ref{fig:circ_dsp}) to quantum computers can lead to inaccurate results, especially when the CNOT gate noise is much higher than the state-preparation errors.

\subsubsection{Obtaining $\delta_{\rm SP}$ by mitigating gate noise}
Fortunately, there are already practical schemes, such as zero-noise extrapolation (ZNE)~\cite{giurgica2020digital} and probabilistic error cancellation (PEC)~\cite{van2023probabilistic}, that can adequately mitigate the CNOT gate noise. Since we only have one CNOT gate in the characterization circuits, the accuracy of the results after mitigation is typically below the state-preparation errors, thus one can still obtain relatively accurate $\delta_{\rm SP}$'s. We will discuss the effects and additional errors on our $\delta_{\rm SP}$-characterization scheme from the gate-noise mitigation methods with more details in \ref{sec:ReForChara}.

Therefore, the $\delta_{\rm SP}$-characterization scheme can still work on noisy devices with an additional gate-noise mitigation procedure, although such mitigation procedure typically requires additional computational resources. Given the ZNE and PEC  procedures have been incorporated in Qiskit~\cite{Qiskit}, below we will present the results on real devices using these mitigation schemes. In the next section, we will introduce a new technique that can efficiently mitigate the CNOT gate noise without additional computational resources. The technique only works for specific circuits on noisy devices with symmetric error channels for Pauli X and Z types. Most noise in current superconducting qubits is dominated by depolarizing noise~\cite{urbanek2021mitigating}, and thus the symmetric condition approximately holds in most superconducting quantum computers. But in any case, one can twirl the CNOT gate so that its error channel becomes a Pauli channel.

\newsavebox\tmpbox

\begin{table*}[t]

\sbox\tmpbox{
\begin{tabular}[b]{|c|c|c|c|c|c|c|c|}
\hline
 $q_t(q_a)$ & $q_0(q_1)$ & $q_1(q_2)$ & $q_2(q_1)$ & $q_3(q_5)$ & $q_4(q_5)$ & $q_5(q_6)$ & $q_6(q_5)$\\ \hline
$\delta_{\rm SPAM}^0$&0.0108(4) &0.0134(5) &0.0086(4) &0.0087(4)&0.0120(4)&0.0162(5)&0.0086(4)\\ 
$\delta_{\rm SPAM}^1$&0.0514(9) &0.0394(8) &0.085(1)  &0.0380(8)&0.0457(8)&0.102(1) &0.0319(7) \\\hline
$\delta_{M}^0$       &0.0005(8) &0.0037(8) &0.0018(9) &0.0020(8)&0.0039(8)&0.010(1) &0.0021(7) \\
$\delta_{M}^1$       &0.0411(8) &0.0297(8) &0.0780(9) &0.0312(8)&0.0376(8)&0.096(1) &0.0253(7) \\\hline
$\delta_{\rm SP}$    &0.011(1)  &0.0101(4) &0.0074(3) &0.0070(5)&0.0085(7)&0.0069(3)&0.0067(4) \\
\hline
\end{tabular} 
}
\renewcommand*{\arraystretch}{0}
  \begin{tabular*}{\linewidth}{@{\extracolsep\fill}p{\wd\tmpbox}p{40mm}@{}}
    \usebox\tmpbox &
    \includegraphics[width=0.16\textwidth]{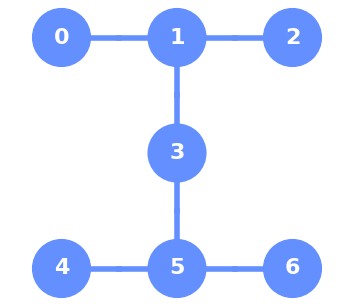} 
    \end{tabular*}

\caption{\label{tb:DspNRB}  The state preparation error and measurement error for all the qubits in ibm\_nairobi  have been characterized using methods described in the text. The qubit connection of ibm\_nairobi is shown on the right. All of the circuits are repeated $6\times 10^5$ times.}
\end{table*}

\subsubsection{Efficient mitigation scheme for symmetric CNOT noise}
In this section, we show that the gate-noise mitigation procedure can be greatly simplified if the noise can be described by uncorrelated single-qubit Pauli noise channels and with some symmetry. We remark that for sufficiently large correlated errors, general gate error mitigation (e.g., ZNE and PEC) may be a more practical approach.

Suppose the noise of the CNOT can be described by a noise channel $\Lambda$ after the CNOT gate, which we assume is a direct product of two Pauli-like single-qubit channels
\begin{equation}
\Lambda(\rho)=\sum_{C \in \{I,X,Y,Z\}} p_C C(\rho) C.
\end{equation}
By adding a random Pauli Z operator with 50\% probability for each qubit before the CNOT gate in Fig.~\ref{fig:circ_dsp}, we ensure that the state-preparation error produces effectively $\rho=  (1-\delta_{\rm SP} )|0\rangle\langle0|+ \delta_{\rm SP}|1\rangle\langle1|$, on which the Pauli Z channel has no effect but other Pauli channels act effectively like the bit-flip error, whose error assignment matrix $A_{\rm CNOT}$ can be combined with the measurement error
\begin{equation}
    A_{\rm M}^{\rm noisy}\equiv A_{M}A_{\rm CNOT},
\end{equation}
where the superscript `noisy' is used to remind us of the noisy CNOT gate. For $q_t$ in Fig.~\ref{fig:circ_dsp}. the SPAM errors become
\begin{equation}
    A_{\rm SPAM}^{\rm noisy}\equiv A_{\rm M}^{\rm noisy} A_{\rm SP}.
\end{equation}

If we assume the CNOT gate noise acts symmetrically on control and target qubits, the $A_{M}^{\rm noisy}$ matrix will not change when swapping the control qubit and target qubit of the CNOT gate. This can be justified by adding additional four noiseless Hadamard gates alongside the CNOT gate to swap the control and target qubits. Then, the error channel will change accordingly $p_x \leftrightarrow p_z$. Thus, if $p_x\approx p_z$ (e.g., when the single-qubit depolarizing noise dominates), the assumption approximately holds (see Appendix.~\ref{app:moreCNOT} for mitigating procedures for more general uncorrelated Pauli noise channels). In this case, one can use the SPAM error with gate noise to calculate $\delta_{\rm SP}$. The SPAM error directly measured from circuits in Fig.~\ref{fig:circ_dsp}, using $q_a$ as the {\it target} of the CNOT gate, becomes 
\begin{gather}\label{eq:dspnoisy}
    \tilde{A}_{\rm SPAM}^{a,\rm noisy}=A_{\rm SPAM}^{a,\rm noisy}A^t_{\rm SP},
\end{gather}
where $A_{\rm SPAM}^{a,\rm noisy}$ can be obtained by swapping CNOT and setting $q_a$ as the {\it control} qubit of the CNOT gate. If one wants to characterize $\delta_{\rm SP}$ for both qubits, one does not need additional circuits, as the circuits after swapping CNOT are exactly $\delta_{\rm SP}$-characterization circuits for $q_a$. Once obtaining $A_{\rm SPAM}^{a,\rm noisy}$ and $\tilde{A}_{\rm SPAM}^{a,\rm noisy}$, we can directly solve for $\delta_{\rm SP}$. As a byproduct, one can also extract the CNOT error rate $\delta_{\rm CNOT}$ and obtain a rough estimation for the rate of the bit-flip channel.

In practice, the procedure of obtaining $\delta_{\rm SP}$ of this method is almost the same as the noiseless case in Eq.~\eqref{eq:spamT}. For general Pauli noise of CNOT gates, the procedure to obtain $\delta_{\rm SP}$ is more complicated, and we leave the discussion in the appendix~\ref{app:moreCNOT}. After obtaining the $\delta_{\rm SP}$, we use the SPAM error without the CNOT gate noise to calculate $\delta_M$ according to Eqs.~\eqref{eq:SPAM=SP+M}. This can be obtained by directly preparing each qubit at $\ket{0/1}$ and then measuring it.

To validate the above arguments, we implement the circuit shown in Fig.~\ref{fig:circ_dsp} on IBM's superconducting quantum computers, and obtain the $\delta_{\rm SP}$ via our efficient mitigation scheme, along with ZNE and PEC. The $\delta_{\rm SP}$ result for $q_{33}$ of ibm\_nazca quantum computer obtained by our efficient mitigation scheme is $\delta_{\rm SP}^{(33)}=0.011(2)$. The results from ZNE and PEC are $0.013(2)$ and $0.014(6)$, respectively, which are consistent with our mitigation schemes. Thus, using our efficient CNOT-noise mitigation scheme merely brings a constant overhead, but can achieve similar accuracy as other gate-noise mitigation schemes. We also add more detailed comparison for our mitigation method and ZNE method in Appendix.~\ref{app:CNOT}.

\subsection{Cloud experimental demonstration}

\subsubsection{Parallel $\delta_{\rm SP}$-characterization for all qubits}
Since our $\delta_{\rm SP}$-characterization only requires two-qubit gates, the mitigation for multiple qubits can be done parallelly. For example, in a 1-D qubit chain, one can do the mitigation for even sites at the same time, then for odd sites.

We use the above method to characterize $\delta_{\rm SP}$ for all qubits in the ibm\_nairobi, and the results are shown in Table~\ref{tb:DspNRB}. We find that the state-preparation error, if not separated from the readout error, is lumped in the SPAM error that is used in the current standard approach for readout mitigation.

\subsubsection{One-qubit state tomography}\label{sec:tomography}
To fully characterize the state-preparation noise channel and confirm the validness of our $\delta_{\rm SP}$-characterization scheme, we also perform state tomography for $\ket{0}$ state and $\ket{1}$ state after obtaining the $\delta_{\rm SP}$ and $\delta_{M}$'s. We perform the experiment on $q_{12}$ of ibmq\_mumbai and get $\delta_{\rm SP}=0.018(1)$. After the measurement mitigation with the characterized $\delta_{M}$'s, the tomography results for preparing $\ket{0}$ and $\ket{1}$ are
\begin{gather}
    \rho_0=\begin{pmatrix}
        0.983(2) &	-0.003(2)-0.016(2)i \\
        -0.003(2)+0.016(2)i &	0.017(2)
    \end{pmatrix},\\
    \rho_1=\begin{pmatrix}
        0.018(2) & 0.002(2) +0.019(2) i \\
        0.002(2) -0.019(2) i &	0.982(2) 
    \end{pmatrix},
\end{gather}
where the diagonal part is consistent with the $\delta_{\rm SP}$ obtained from our quantification scheme.

\subsection{Resource estimations}
\label{sec:ReForChara}

In this section, we estimate the resource requirements for our $\delta_{\rm SP}$-characterization scheme, especially the measurement number and circuit number requirements, where we count each full-qubit measurement of one circuit as one measurement.

Firstly, $Z$-twirling can be implemented by four circuits: with/without the initial $Z$ gate for the target and the ancilla qubits. The probability distribution for $50\%$-Z twirling can be obtained by directly averaging the results of the four circuits, which only brings a constant factor of uncertainty that can be canceled by increasing the number of measurements. Thus, the total overhead from $Z$-twirling is a constant number. In addition, our $\delta_{\rm SP}$-characterization scheme can be performed in parallel for multiple pair of qubits. Thus, the total number of measurements will not grow with the number of qubits due to the parallelization, although the complexity for classical data processing grows linearly.

In the case of noiseless CNOT, the error of our $\delta_{\rm SP}$-characterization is purely statistical uncertainty, which can be arbitrarily small by increasing the number of measurements $N$, with the corresponding uncertainty scale as $O(\frac{1}{\sqrt{N}})$. 

In the noisy CNOT case, the accuracy of our scheme is upper bounded by the accuracy of gate-noise mitigation. We assume after some gate-noise mitigation scheme with a total of $N_{G}$ measurements of circuits, we can obtain expectations for Z operators with systematic error $\Delta_{G}$ and variance $\sigma_G^2$. As the systematic error of the $\delta_{\rm SP}$ is lower bounded by $\Delta_{G}$, our $\delta_{\rm SP}$-characterization scheme only works if $\Delta_G \ll \delta_{\rm SP}$ and $\sigma_G$ is sufficiently small $\sigma_G\ll \delta_{\rm SP}$. Supposing the $\Delta_{G}$ can be made small compared to $\delta_{\rm SP}$, after performing $N_{\rm SP}$ repetitions of gate-noise-mitigated circuits (each repetition contains $N_G$ measurements), the deviation $\Delta(\delta_{\rm SP})$ from the true value of $\delta_{\rm SP}$ is bounded, with probability $1-p$,
\begin{equation}
    |\Delta(\delta_{\rm SP})|< C|\Delta_{G}|+O\left(\sigma_G\sqrt{\frac{\log\frac{1}{p}}{N_{\rm SP}}}\right),
\end{equation}
where $C>0$ is a constant related to the true value of state-preparation errors and measurement errors. If all $\delta_{\rm SPAM}$'s are small, $C$ is close to 2.

In typical gate-noise mitigation schemes (e.g., ZNE and PEC), $\sigma_G$  roughly scales like $O(\frac{1}{\sqrt{N_G}})$. Thus, the total uncertainty still scales as $O(\frac{1}{\sqrt{N}})$, where $N=N_{\rm SP} N_G$. Therefore, the uncertainty of our $\delta_{\rm SP}$-characterization is directly related to the uncertainty of the chosen gate-noise mitigation $\sigma_G$, and we need $N_{\rm SP}$ repetitions of those mitigated results (with total measurement number $N=N_{\rm SP} N_G$) to further reduce the uncertainty from $O(\sigma_G)$ by a factor $\frac{1}{\sqrt{N_{\rm SP}}}$.

Nevertheless, the systematic error of $\delta_{\rm SP}$ is still lower-bounded by those of gate-noise mitigation schemes. Therefore, we require that the systematic error of gate-noise mitigation is much smaller than the $\delta_{\rm SP}$ in order to obtain meaningful results. As we use only one CNOT in the characterization circuits, in current noisy superconducting quantum computers where CNOT gate errors can be small, the condition is roughly satisfied for typical gate-noise mitigation schemes, such as ZNE and PEC.

\section{Complete mitigation for measurement error and state-preparation error}
\label{sec:completemitigation}

Measurement error mitigation is an essential tool to extract information from noisy quantum devices. The measurement error with only bit flips can be characterized by the error assignment matrix. In the single-qubit case, the matrix is simply 
\begin{eqnarray}\label{eq:am}
    A_M=\begin{pmatrix}
        1-\delta_M^0 & \delta_M^1\\
        \delta_M^0 & 1-\delta_M^1
    \end{pmatrix}.
\end{eqnarray}
Note that we assume that there are only bit-flip errors, or else X and Y errors have been removed by the Z randomization procedure mentioned above.
Assume the ideal preparation of the initial state, the effect of the measurement error on the ideal outcome probability distribution $\mathcal{P}$ is then
\begin{equation}
    \bm{P}_{M}=A_{M}\mathcal{P},
\end{equation}
where $A_{M}$ can be the general assignment matrix characterizing $n$-qubit measurement error and $\bm{P}_{M}$ denotes the noisy probability distribution affected by only measurement errors. Suppose we know the exact form of $A_M$; we can exactly mitigate the measurement error by
\begin{equation}
\label{eq:AmMiti}
    \mathcal{P}=A_M^{-1}\bm{P}_{M}.
\end{equation}
However, when the state preparation error appears, the matrix $A_M$ is usually hard to obtain exactly.
In practice, the state preparation and readout errors are usually lumped together as the SPAM error $\delta_{\rm SPAM}$. The corresponding assignment matrix $A_{\rm SPAM}$ is obtained by naive calibration circuits, such as in Qiskit~\cite{Qiskit}, where $2^n$ computational-basis states are prepared and then measured. Subsequently, the inverse of the $A_{\rm SPAM}$, instead of $A_M$, is applied for the mitigation in practice
\begin{equation}
    \hat{\bm{P}}=A_{\rm SPAM}^{-1}\bm{P}_{\rm SP,M}=A_{\rm SPAM}^{-1} A_M \bm{P}_{\rm SP}.
\end{equation}
where $\bm{P}_{\rm SP,M}$ denotes the noisy probability distribution affected by both state preparation and measurement errors, and $\bm{P}_{\rm SP}$ denotes the noisy probability distribution affected by only state-preparation errors. Since $A_{\rm SPAM}^{-1} A_M$ is not identity when $\delta_{\rm SP}\neq 0$, the mitigated result $\hat{\bm{P}}$  differs from the $\delta_{\rm SP}$-only noisy distribution $\bm{P}_{\rm SP}$. Suppose the assignment matrix for the state-preparation error on the initial state is $A_{\rm SP}$. In single-qubit case, it is 
\begin{eqnarray}\label{eq:ASP}
    A_{\rm SP}=\begin{pmatrix}
        1-\delta_{\rm SP} & \delta_{\rm SP}\\
        \delta_{\rm SP} & 1-\delta_{\rm SP}
    \end{pmatrix}.
\end{eqnarray}

What calibration circuits measure is effectively the successive application of $A_{\rm SP}$ and $A_{M}$ on the initial state, which gives $A_{\rm SPAM}=A_M A_{\rm SP}$. Thus, the result mitigated by $A_{\rm SPAM}$ is
\begin{equation}
    \hat{\bm{P}}=A_{\rm SP}^{-1} \bm{P}_{\rm SP}.
\end{equation}
Note that the distribution $\bm{P}_{\rm SP}$ contains the state-preparation error, which takes place at the beginning of the circuit and, in general, cannot be mitigated by applying the inverse at the end of the circuit. Thus, the mitigation by $A_{\rm SPAM}^{-1}$ usually over-mitigates the outcome and can give rise to some unphysical probability distributions (in addition to those caused by statistical fluctuations), such as negative probabilities. Nevertheless, the negative probability problem can be dealt with by using the nearest physical probability distribution instead~\cite{smolin2012efficient}, after applying the mitigation matrix $A_{\rm SPAM}^{-1}$.

While fully characterizing those error matrices requires exponentially many circuits, if we assume all the errors are local, those error assignment matrices can be decomposed into a direct product of all single-qubit measurement errors $\{A^{(i)}\}$
\begin{gather}
\label{eq:ami}
    A_{M}=A_M^{(1)}\otimes A_M^{(2)}\cdots A_M^{(n)},\\
    \label{eq:aspi}
    A_{\rm SP}=A_{\rm SP}^{(1)}\otimes A_{\rm SP}^{(2)}\cdots A_{\rm SP}^{(n)},\\
    \label{eq:aspami}
    A_{\rm SPAM}=A_{\rm SPAM}^{(1)}\otimes A_{\rm SPAM}^{(2)}\cdots A_{\rm SPAM}^{(n)},
\end{gather}
where
\begin{eqnarray}
    A_{\rm SPAM}^{(i)}=\begin{pmatrix}
        1-\delta_{\rm SPAM}^{i,0} & \delta_{\rm SPAM}^{i,1}\\
        \delta_{\rm SPAM}^{i,0} & 1-\delta_{\rm SPAM}^{i,1}
    \end{pmatrix}.
\end{eqnarray}
The above conditions approximately hold and match many current superconducting qubit  devices~\cite{nation2021scalable}, where the uncorrelated errors dominate, and correlated errors among qubits are usually small. It is easy to check that $A_{\rm SPAM}^{(i)}=A_M^{(i)} A_{\rm SP}^{(i)}$. Thus, one can measure the $\delta_{\rm SPAM}$ locally and then combine them to get the full $A_{\rm SPAM}$ matrix, which only takes $O(n)$ number of circuits. Throughout our work, we will mainly focus on uncorrelated errors so that all error assignment matrices can be decomposed into a tensor product of single-qubit ones.

Nevertheless, the state-preparation error usually does not commute with the circuit gates, which, in principle, cannot be automatically combined with the measurement errors at the end of the circuit. Thus, we need a new technique to mitigate the state-preparation error and the measurement error separately.

\subsection{Measurement error mitigation}
In this section, we present a similar mitigation scheme for the measurement error as Eq.~\ref{eq:AmMiti}. Here, we assume that all the errors are uncorrelated and Eqs.~\eqref{eq:ami}, \eqref{eq:aspi}, and \eqref{eq:aspami} hold. We can firstly measure the SPAM error $\delta_{\rm SPAM}$ for each qubit, then use the methods in Sec.~\ref{sec:dsp} to obtain the state-preparation errors $\delta_{\rm SP}$'s and measurement errors $\delta_{M}$'s. After getting the single-qubit measurement errors $\delta_{M}^{i,0}$ and $\delta_{M}^{i,1}$ for each qubit, we construct the full measurement error assignment matrix $A_M$ according to Eq.~\eqref{eq:am} and Eq.~\eqref{eq:ami}. The measurement error is mitigated by applying the inverse of $A_M$
\begin{equation}
\label{eq:Mdm}
\hat{\bm{P}}_{\rm SP}=A_M^{-1} \bm{P}_{\rm SP,M}.
\end{equation}
Ideally, after such a readout error mitigation process, the remaining errors in $\hat{\bm{P}}_{\rm SP}$ purely come from state preparation and gate operations. 

\subsection{State preparation error mitigation}

\begin{figure}[h]
    \centering
\qquad\qquad$\left.\begin{tabular}{l}
    \Qcircuit @C=0.1em @R=.1em @! {
\lstick{\ket{0}_1^{\rm Noisy}}&  \multigate{3}{~U~} & \meter  \\
\lstick{\ket{0}_2^{\rm Noisy}}&  \ghost{~U~}        & \meter  \\
\lstick{\vdots}   &  \nghost{~U~}        & \vdots  \\
\lstick{\ket{0}_n^{\rm Noisy}}&  \ghost{~U~}        & \meter  }
\end{tabular}~\right\}\bm{P}_{\rm raw}$
    \caption{The target circuit to be mitigated, where each qubit is intended to be initialized at $\ket{0}$, but with some probabilities to be prepared at $\ket{1}$ due to the state-preparation errors. We label such erroneous preparation as $\ket{0}^{\rm Noisy}$. The gates $U$ and measurements can also contain noise. The output probability distribution is labeled as $\bm{P}_{\rm raw}$. }
    \label{fig:circ_P}
\end{figure}

In this section, we present a new mitigation scheme to mitigate the state-preparation errors. The mitigation is applied in a fully noisy scenario: we prepare all $n$ qubits at $\ket{0}$ with state-preparation errors, apply some noisy gates $U$, perform some noisy measurements at the end, and obtain a noisy probability distribution $\bm{P}_{\rm raw}$ (see Fig.~\ref{fig:circ_P}). The goal of our state-preparation error mitigation is to get a mitigated distribution $\hat{\bm{P}}$ that is sufficiently close to the  $\delta_{\rm SP}$-free distribution $\mathcal{P}$, where $\mathcal{P}$ can be obtained with perfect qubit initialization at all $\ket{0}$'s along with the same noisy gates $U$ and noisy measurements as $\bm{P}_{\rm raw}$.

As we assume the state-preparation errors can be described by independent classical bit-flip errors, the erroneous output can be regarded as samples of classical independent random events. For each event, qubits $q_j$'s are perfectly initialized in $s_j\in\{0,1\}$, undergo the same noisy gates $U$ and noisy measurements, and output a probability distribution $\mathcal{P}_{s_1,s_2...s_n}$. Such event happens with a probability $\prod_{j=1}^n q_{\rm SP}^{(j)}(s_j,0) $, where 
\begin{eqnarray}\label{eq:qSP}
    q_{\rm SP}^{(j)}(s,s') \equiv 
        \begin{cases}
            1-\delta_{\rm SP}^{(j)},& s=s';\\
            \delta_{\rm SP}^{(j)}, & s\neq s';
        \end{cases}
\end{eqnarray}
and $\delta_{\rm SP}^{(j)}$ is the state-preparation error of the $j$-th qubit $q_j$. Thus, the output probability distribution $\bm{P}$ is a combination of all possible initial assignments $\{s_j\}\in \{0,1\}^{n}$ with the corresponding probability
\begin{equation}\label{eq:pdsp}
    \bm{P}_{\rm raw}=\sum_{\{s_j\}\in \{0,1\}^{n}} \left(\prod_{j=1}^n q_{\rm SP}^{(j)}(s_j,0) \right)\mathcal{P}_{s_1,s_2...s_n}.
\end{equation}
We can also regard the $\bm{P}_{\rm raw}$ as a function of $\delta_{\rm SP}^{(j)}$'s, and rewrite the equation as
\begin{equation}\label{eq:px1x2}
    \bm{P}(x_1,x_2,...)=\sum_{\{s_j\}\in \{0,1\}^{n}} \left(\prod_{j=1}^n q_{x_j}(s_j,0) \right)\mathcal{P}_{s_1,s_2...s_n},
\end{equation}
where $q_{x_j}(s,s')$ is $1-x_j$ if $s=s'$ and $x_j$ otherwise, similar to Eq.~\eqref{eq:qSP}. $\bm{P}_{\rm raw}$ can then be obtained from $\bm{P}(x_1,x_2,...)$ by setting $\{x_j=\delta_{\rm SP}^{(j)}\}$.

Note that the Eq.~\eqref{eq:px1x2} is a linear equation with regard to $\mathcal{P}_{s_1,s_2...s_n}$'s. Thus, to obtain desired $\delta_{\rm SP}$-free probability distribution
\begin{equation}
   \mathcal{P}=\mathcal{P}_{0,0...}= \bm{P}(0,0...),
\end{equation}
we can, in principle, solve $2^n$ linear equations with $2^n$ choices of $\{x_1,x_2,...\}$ for $\bm{P}(x_1,x_2,...)$ in Eq.~\eqref{eq:px1x2}. Nevertheless, in real quantum devices, it is usually hard to tune the state-preparation error freely. Here we introduce a trick to change $x_j$'s without changing the actual state-preparation errors. Notice that if we add a noiseless initial X gate on qubit $q_j$, then the qubit is prepared in $\ket{0}$ with probability $\delta_{\rm SP}^{j}$, and in $\ket{1}$ with probability $1-\delta_{\rm SP}^{j}$, which effectively changes the state-preparation error for $q_j$ to $1-\delta_{\rm SP}^{j}$. Thus, as long as $q_j\neq\frac{1}{2}$, we can obtain a linearly independent equation for $\mathcal{P}_{s_1,s_2...s_n}$'s. By adding or not adding initial X gates for all qubits, we obtain $2^n$ configuration of $\{x_j\}$'s , and $2^n$ linear equations
\begin{equation}\label{eq:dspEqs}
    \bm{P}^{s_1',s_2',...s_n'}=\sum_{\{s_j\}\in \{0,1\}^{n}} \left(\prod_{j=1}^n q_{\rm SP}^{(j)}(s_j,s_j') \right)\mathcal{P}_{s_1,s_2...s_n},
\end{equation}
where $s_j,s_j'\in \{0,1\}$ indicate qubit-initialization choices and
\begin{multline}
    \bm{P}^{s_1',s_2',...s_n'}\equiv\bm{P}\left(s_1'(1-\delta_{\rm SP}^{(1)})+(1-s_1')\delta_{\rm SP}^{(1)},\right.\\
    \left. s_2'(1-\delta_{\rm SP}^{(2)})+(1-s_2')\delta_{\rm SP}^{(2)},...\right).
\end{multline}
The probability distribution $\bm{P}^{s_1',s_2',...s_n'}$ can be experimentally obtained, by initializing the qubits in $\ket{s_1',s_2',...s_n'}$ via adding an initial X gate to qubit $q_j$ if $s_j'=1$. For a small number of qubits, one can exactly solve $\mathcal{P}$'s  by explicitly measuring the $2^n$ circuits and solving $2^n$ linear equations.

However, for a large number of qubits, exactly solving $\mathcal{P}$ requires exponentially many circuits and is not practical in experiments. Here, we present an approximate approach that takes a linear number of circuits to achieve $O(\delta_{\rm SP}^2)$ accuracy.
To do this, we first expand the distribution $\bm{P}$ near the $\delta_{\rm SP}^{(j)}$,
\begin{multline}\label{eq:Pexpansion}
    \bm{P}(x_1,x_2...)\approx\bm{P}(\delta_{\rm SP}^{(1)},\delta_{\rm SP}^{(2)},...)+\\
    \sum_{i=1}^n \frac{\partial\bm{P}(\delta_{\rm SP}^{(1)},...)}{\partial x_i} (x_i-\delta_{\rm SP}^{(j)})+...
\end{multline}
If we assume the $\delta_{\rm SP}^{(i)}$'s are small and approximate the above equation to the order of $O(\delta_{\rm SP})$, we can neglect higher order terms and get the first-order estimation $\hat{\bm{P}}$ of $\bm{P}(0,0...)$
\begin{equation}\label{eq:dspm0}
    \hat{\bm{P}}=\bm{P}(\delta_{\rm SP}^{(1)},\delta_{\rm SP}^{(2)}...)-\sum_{i=1}^n \frac{\partial\bm{P}(\delta_{\rm SP}^{(1)},...)}{\partial x_i} \delta_{\rm SP}^{(i)},
\end{equation}
where $\bm{P}(\delta_{\rm SP}^{(1)},\delta_{\rm SP}^{(2)}...)=\bm{P}_{\rm raw}$ is the raw experimental result to be mitigated, and the derivatives $\frac{\partial\bm{P}(\delta_{\rm SP}^{(1)},...)}{\partial x_i}$ can be obtained in a similar way as solving linear equations. This can be seen by rewriting Eq.~\eqref{eq:px1x2} into
\begin{equation}\label{eq:p1}
    \bm{P}(x_1,\delta_{\rm SP}^{(2)},...)=(1-x_1)\tilde{\bm{P}}_{0,X}+x_1 \tilde{\bm{P}}_{1,X},
\end{equation}
where we have taken $q_1$ as an example and
\begin{eqnarray}\nonumber
    \tilde{\bm{P}}_{0,X}&=&\sum_{\{s_2,s_3...\}\in \{0,1\}^{n-1}} \left(\prod_{j=2}^n q_{\rm SP}^{(j)}(s_j,0) \right)\mathcal{P}_{0,s_2...s_n},\\\nonumber
    \tilde{\bm{P}}_{1,X}&=&\sum_{\{s_2,s_3...\}\in \{0,1\}^{n-1}} \left(\prod_{j=2}^n q_{\rm SP}^{(j)}(s_j,0) \right)\mathcal{P}_{1,s_2...s_n}.
\end{eqnarray}
Thus, $\bm{P}(x_1,\delta_{\rm SP}^{(2)},...)$ is linear with $x_1$. The derivative can be obtained by one additional probability distribution with $x_1\neq \delta_{\rm SP}^{(1)}$, and such different $x_1$ can be obtained by adding initial X gate on $q_1$ so that effectively $x_1=1- \delta_{\rm SP}^{(1)}$. We denote the probability distribution obtained by adding an initial noiseless X gate on $q_i$ as $\bm{Q}_i$ (see Fig.~\ref{fig:circ_Qi}).

\begin{figure}[h]
    \centering
\qquad\qquad$\left.\begin{tabular}{l}
    \Qcircuit @C=0.1em @R=.1em @! {
\lstick{\ket{0}_1^{\rm Noisy}}& \qw    &  \multigate{4}{~U~} & \meter  \\
\lstick{\vdots}               &        & \nghost{~U~}        & \vdots  \\
\lstick{\ket{0}_i^{\rm Noisy}}&\gate{X}& \ghost{~U~}         & \meter  \\
\lstick{\vdots}               &        &  \nghost{~U~}       & \vdots  \\
\lstick{\ket{0}_n^{\rm Noisy}}& \qw    &  \ghost{~U~}        & \meter  }
\end{tabular}~\right\}\bm{Q}_{i}$
    \caption{The first-order state-preparation error mitigation circuit on $i$-th qubit $q_i$, where an initial X gate is added to the $q_i$ and we assume the noise brought by this additional X is negligible. The rest of the circuits, including  the noisy qubit preparation, noisy gates $U$, and noisy measurements are the same as target circuit in Fig.~\ref{fig:circ_P}. The output probability distribution is labeled as $\bm{Q}_{i}$. }
    \label{fig:circ_Qi}
\end{figure}

The original distribution $\bm{P}_{\rm raw}$ and the additional distribution $\bm{Q}_1$ (the distribution from the same circuit with additional initial X gate on the first qubit) can be viewed as the result of substituting $x_1=\delta_{\rm SP}^{(1)}$ and $x_1=1-\delta_{\rm SP}^{(1)}$, respectively, in Eq.~\eqref{eq:p1},
\begin{eqnarray}\nonumber
    \bm{P}_{\rm raw}&\equiv&\bm{P}(\delta_{\rm SP}^{(1)},...)=(1-\delta_{\rm SP}^{(1)})\tilde{\bm{P}}_{0,X}+\delta_{\rm SP}^{(1)} \tilde{\bm{P}}_{1,X},\\\nonumber
    \bm{Q}_1&\equiv&\bm{P}(1-\delta_{\rm SP}^{(1)},...)=\delta_{\rm SP}^{(1)}\tilde{\bm{P}}_{0,X}+(1-\delta_{\rm SP}^{(1)}) \tilde{\bm{P}}_{1,X}.
\end{eqnarray}
Thus, we can obtain the derivative of $x_1$ by
\begin{equation}\label{eq:dp1}
\begin{aligned}
    \frac{\partial\bm{P}(\delta_{\rm SP}^{(1)},...)}{\partial x_1}=&\tilde{\bm{P}}_{1,X}-\tilde{\bm{P}}_{0,X}\\
    =&\frac{\bm{Q}_1-\bm{P}_{\rm raw}}{1-2 \delta_{\rm SP}^{(1)}}.
\end{aligned}
\end{equation}
The other derivatives w.r.t.  $x_i$'s can be obtained in a similar way. Substituting them into the Eq.~\eqref{eq:dspm0}, we arrive at the equation for the first-order state-preparation error mitigation
\begin{equation}\label{eq:Mdsp}
    \hat{\bm{P}}=\bm{P}_{\rm raw}+\sum_{i=1}^n \frac{\delta_{\rm SP}^{(i)}}{1-2 \delta_{\rm SP}^{(i)}}(\bm{P}_{\rm raw}-\bm{Q}_i) .
\end{equation}
Note that the state-preparation error mitigation \eqref{eq:Mdsp} is linear among different distributions $\bm{P}$'s and that measurement mitigation \eqref{eq:Mdm} is a linear transformation within each distribution $\bm{P}$, so swapping the order of measurement error mitigation and state-preparation error mitigation here does not change the final results.

We remark that one can achieve $r$-th-order accuracy mitigation at the cost of $O(n^r)$ additional circuits with a similar procedure, where $n$ is the qubit number. A straightforward way is to expand the Eq.~\eqref{eq:Pexpansion} to higher orders. As $P(x_1,x_2...)$ is a multi-linear function of $\{x_1,x_2...\}$, a $r$-th order derivative is non-zero only if it involves exactly $r$ number of $x_j$'s. To obtain the $r$-th order derivative with respect to $x_{m_1},...,x_{m_r}$, one can follow the procedure similar to those from Eq.~\eqref{eq:p1} to Eq.~\eqref{eq:dp1}. By adding/not adding initial X gate for qubits $\{x_{m_1},...,x_{m_r}\}$, one can obtain $2^r$ number of linear equations with respect to $2^r$ number of $\tilde{\bm{P}}$'s, which can be exactly calculated by solving the $2^r$ linear equations. In addition, the $r$-th order derivative can  be expressed as linear combinations of $\tilde{\bm{P}}$, thus can also be directly calculated after obtaining $\tilde{\bm{P}}$'s. For $r$-th order derivatives, the number of additional distributions needed is $\sum_{j=1}^r {n \choose j}$, and these distributions can be obtained by adding initial X gates for at most $r$ qubits.

Since the $\delta_{\rm SP}$-mitigation only involves the linear combination of probability distributions, we can directly apply the mitigation scheme for observables. For an observable $O_{\rm raw}$ obtained from a noisy quantum computer, we can perform the circuits with initially flipped $i$-th qubit to obtain $O_i$, and mitigate (at the first order) the state-preparation errors by
\begin{equation}
    \hat{O}=O_{\rm raw}+\sum_{i=1}^n \frac{\delta_{\rm SP}^{(i)}}{1-2 \delta_{\rm SP}^{(i)}}(O_{\rm raw}-O_i).
\end{equation}

In addition, as the $\delta_{\rm SP}$-mitigation scheme does not interfere with each probability distribution, the scheme is compatible with most gate-noise mitigation and measurement error mitigation techniques. To do further mitigation for gate noises and measurement errors, one can firstly apply corresponding mitigation techniques to $\bm{P}_{\rm raw}$ and $\bm{Q}_i$, and then do the  $\delta_{\rm SP}$-mitigation according to Eq.~\eqref{eq:Mdsp}. Same procedure can also be applied to the additional gate and measurement noise mitigations for observables.

\subsection{Resource estimations}
\label{sec:resourceForMiti}

Compared to the results without $\delta_{\rm SP}$-mitigation, our $\delta_{\rm SP}$-mitigation scheme requires additional $n$ circuits, where $n$ is the qubit number. The $\delta_{\rm SP}$-mitigation reduce the error rate to $O(\delta_{\rm SP}^2)$, whereas the error rate is $O(\delta_{\rm SP})$ without mitigation. On the other hand, when doing the $\delta_{\rm SP}$-mitigation according to Eq.~\eqref{eq:Mdsp}, the uncertainty of the final results $\hat{\bm{P}}$ grows with $O(\sqrt{n}\delta_{\rm SP})$, where we assume all state-preparation errors are almost identical and close to some $\delta_{\rm SP}<\frac{1}{3}$ and $n$ is sufficiently large. To remedy this and restore the same uncertainty before $\delta_{\rm SP}$-mitigation, we require additional $O( n \delta_{\rm SP}^2)$ multiples of measurements to reduce the statistical uncertainty. Thus, if one strictly requires the mitigated $\hat{\bm{P}}$ to achieve the same uncertainty as the $\bm{P}_{\rm raw}$, the total additional overhead of measurements is $O(\delta_{\rm SP}^2 n^2)$.

We remark that the overhead can be greatly reduced if not all qubits are highly noisy. In the first order mitigation, the correction terms for $\delta_{\rm SP}$ of each qubit are separated. Thus, one can perform qubit-wise mitigation only for those qubits with large $\delta_{\rm SP}$'s. For example, if only $q_j$ contains large state-preparation error and other qubits are almost perfectly prepared with negligible $\delta_{\rm SP}$'s, one can mitigate the state-preparation errors for $q_j$ only by
\begin{equation}
    \hat{\bm{P}}=\bm{P}_{\rm raw}+ \frac{\delta_{\rm SP}^{(j)}}{1-2 \delta_{\rm SP}^{(j)}}(\bm{P}_{\rm raw}-\bm{Q}_j) ,
\end{equation}
with the remainder error being $O\big((\delta_{\rm SP}^{(j)})^2+\sum_{i\neq j}\delta_{\rm SP}^{(i)}\big) $. Therefore, the total overhead of the mitigation can be reduced to $O(\delta^2 m^2)$, where $m$ counts the number of qubits with sufficiently large $\delta_{\rm SP}\sim \delta$ that need mitigations.

\begin{figure*}[t]
\begin{tabular}{ll}
(a) & (b)  \\
\includegraphics[width=0.45\textwidth]{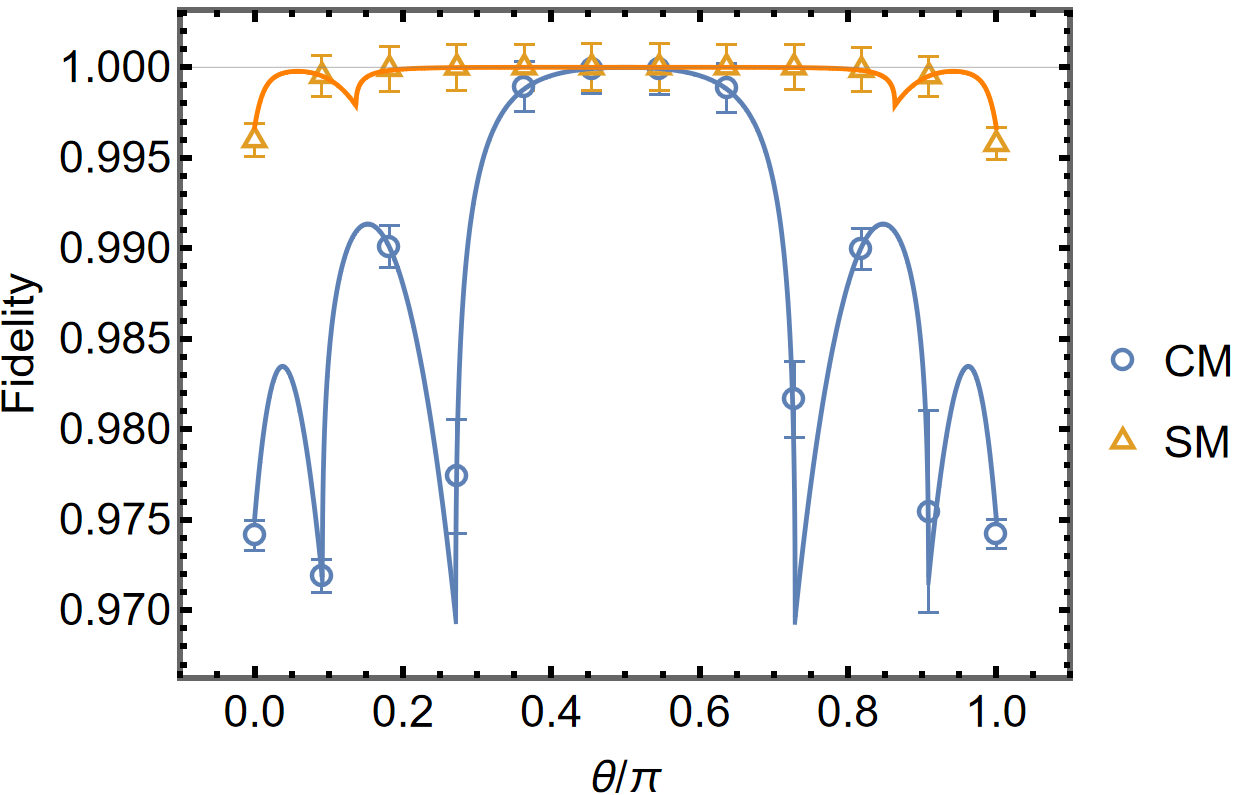}&
\includegraphics[width=0.445\textwidth]{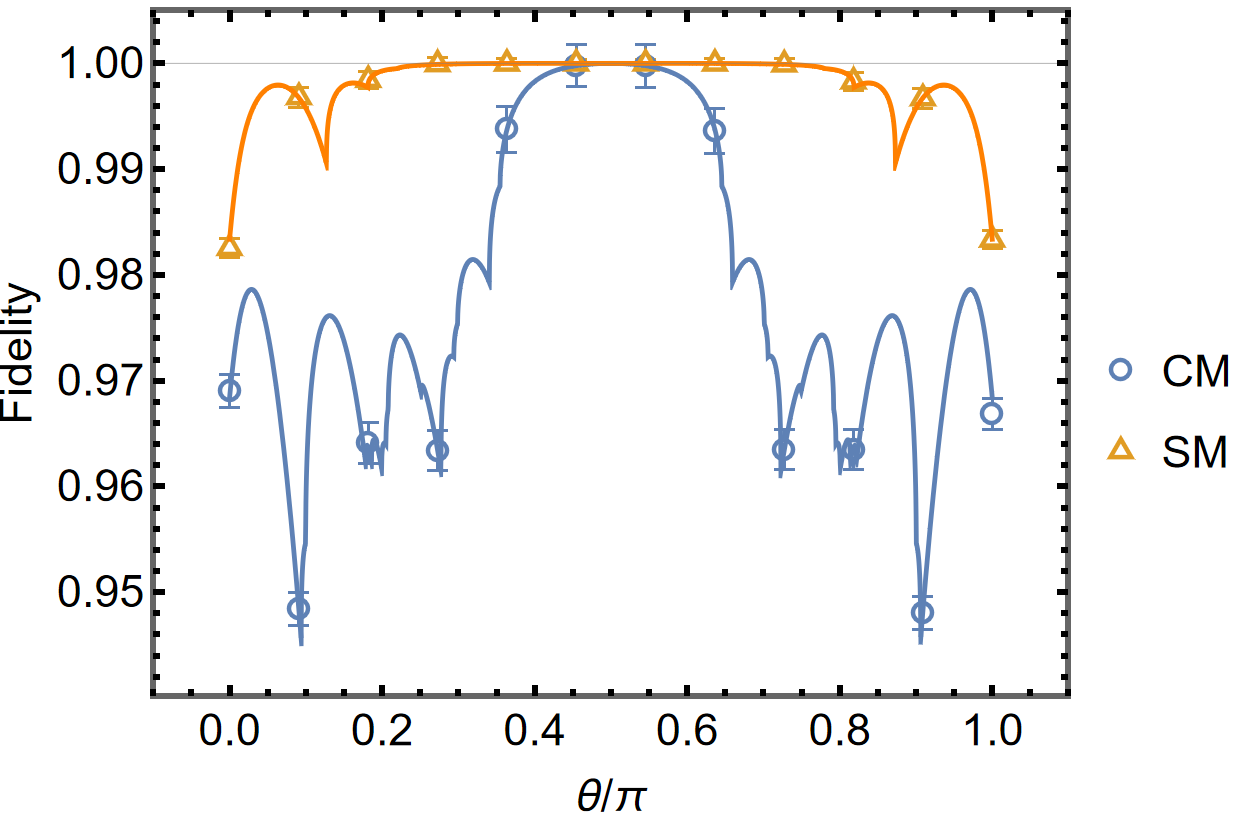}\\
\end{tabular}
    \caption{Simulation results for comparison between two mitigation schemes for the (a) 2-qubit circuit~(Fig.~\ref{fig:circ_sim2q}) and (b) 4-qubit circuit~(Fig.~\ref{fig:circ_sim4q}). We set $\delta_{\rm SP}=0.05$, $\delta_M^0=0.04$, and $\delta_M^1=0.06$ for all qubits. The circuits for calculating $\delta_{\rm SP}$ are repeated $4\times 10^6$ times, and the target circuits are repeated $6.4\times 10^5$ times.}
    \label{fig:dspm_sim}
\end{figure*}

\section{comparison between two mitigation schemes}
\label{sec:comparison}
In the current standard practice on superconducting qubits, readout mitigation is carried out first by measuring the mitigation assignment matrix $A_{\rm SPAM}$ obtained by measuring the distribution in the computational basis when states are prepared in all different computational states. Then, the inverse of the assignment matrix is applied to the actual measurement distribution. However, as mentioned before, the preparation of computational states incurs errors, and these are not separately characterized and are thus lumped together in the readout mitigation matrix. We call this combined mitigation (CM) and refer to ours as separate state preparation and readout mitigation (SM).

\subsection{Summary of the two mitigation schemes}

First, we summarize the procedure of CM as follows:
\begin{quote}
    1. Execute the quantum circuit $U$ on qubits $\{q_i\}$ and get the noisy raw probability distribution $\bm{P}_{\rm raw}$;\\
    2. Run the standard SPAM error calibration circuits (initializing each qubit at $\ket{0/1}$ followed by measurements) on qubits $\{q_i\}$ and get the SPAM error assignment matrix $A_{\rm SPAM}$;\\
    3. Apply the inverse of $A_{\rm SPAM}$ to the raw distribution and get the mitigated quasi-probability distribution $\tilde{\bm{P}}=A_{\rm SPAM}^{-1}\bm{P}_{\rm raw}$;\\
    4. Find the nearest probability distribution $\hat{\bm{P}}$ from $\tilde{\bm{P}}$ in case of the existence of negative probabilities in $\tilde{\bm{P}}$. Then, the output $\hat{\bm{P}}$ is the desired mitigated probability distribution.
\end{quote}
Next, we summarize our proposed mitigation procedure (SM) as follows:
\begin{quote}
    1. Execute the quantum circuit $U$ on qubits $\{q_i\}$ and get the noisy raw probability distribution $\bm{P}_{\rm SP,M}$;\\
    2. Run the calibration circuits in sec.~\ref{sec:dsp} on each qubit pair and separately get the state preparation error $\{\delta_{\rm SP}^{(i)}\}$ and readout error assignment matrix $A_{M}$;\\
    3. Add an initial X gate on $i$th qubit $q_i$ and then apply the same circuit $U$ for each qubit to get the calibration distribution $\{\bm{Q}_i\}$;\\
    4. Perform the measurement error mitigation by applying the inverse of $A_{M}$ to the raw distribution $\bm{P}_{\rm SP,M}$ and calibration distribution $\{\bm{Q}_i\}$s, and get the mitigated quasi-probability distribution $\tilde{\bm{P}}_{\rm SP}=A_{M}^{-1}\bm{P}_{\rm SP,M}$ and $\{\tilde{\bm{Q}}_i=A_{M}^{-1}\bm{Q}_{i}\}$;\\
    5. Use the Eq.~\eqref{eq:Mdsp} to get the $\delta_{\rm SP}$-mitigated quasi-probability distribution $\tilde{\bm{P}}$ from $\tilde{\bm{P}}_{\rm SP}$, $\{\tilde{\bm{Q}}_i\}$, and $\{\delta_{\rm SP}^{(i)}\}$ : $\tilde{\bm{P}}=\tilde{\bm{P}}_{\rm SP}+\sum_{i=1}^n \frac{\delta_{\rm SP}^{(i)}}{1-2 \delta_{\rm SP}^{(i)}}(\tilde{\bm{P}}_{\rm SP}-\tilde{\bm{Q}}_i) $;\\
    6. Find the nearest probability distribution $\hat{\bm{P}}$ from $\tilde{\bm{P}}$ in case of the existence of negative probabilities in $\tilde{\bm{P}}$. Then, the output $\hat{\bm{P}}$ is the desired mitigated probability distribution.
\end{quote}

\begin{figure*}[t]
    \centering
    \includegraphics[width=0.9\textwidth]{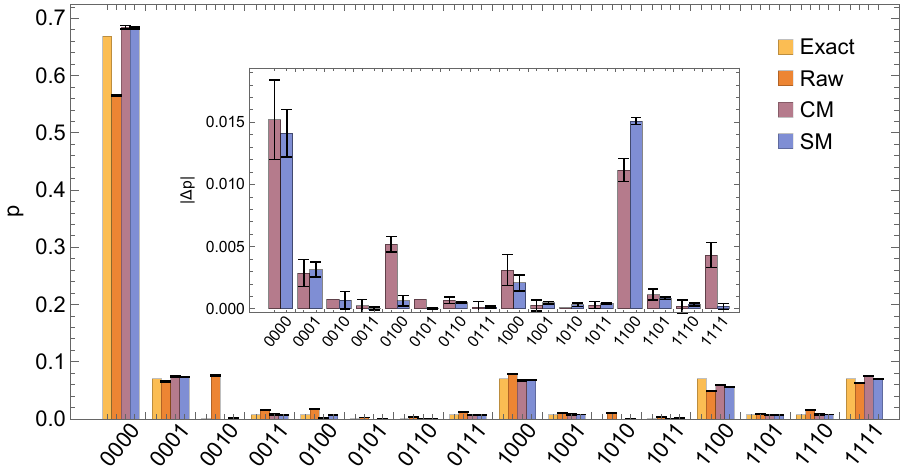}
    \caption{Experimental results, expected exact results, and mitigated results, using both combined mitigation (CM) and separate mitigation (SM), for the 4-qubit circuit shown in Fig.~\ref{fig:circ_sim4q} with $\theta=\pi/5$. The outcome labels on the horizontal axis are in the order $[q_4,q_3,q_2,q_1]$. We also illustrate the absolute differences $|\Delta p|$ from exact values for CM and SM in the inset. The results are obtained from ibmq\_mumbai (whose machine layout is shown above the histogram), with the initial layout of our experiment being the four qubits [21,18,15,12] (highlighted in green in Fig.~\ref{fig:mumbai}). The circuits for calculating $\{\delta_{\rm SP}\}$ are repeated $ 6\times10^5$ times, and the target circuits are repeated $ 10^5$ times.}
    \label{fig:exprmt4q}
\end{figure*}

In both CM and SM, we assume all the state-preparation errors and measurement errors are uncorrelated for different qubits so that their assignment matrices can be written as a tensor product of single-qubit ones. We remark that, the total complexity and resource requirements for CM are the same as obtaining $\tilde{\bm{P}}_{\rm SP}=A_{M}^{-1}\bm{P}_{\rm SP,M}$ in SM. The only difference is that in CM, we use $A_{\rm SPAM}$ to mitigate the measurement errors, whereas in SM, we use $A_{\rm M}$ instead, which is closer to the true measurement errors. Since SM requires additional circuits and measurements for $Q_{i}$'s, it contains an additional overhead for measurement number. According to the analysis in Sec.~\ref{sec:resourceForMiti},
if $\delta_{\rm SP}^2 n\ll 1$, the additional overhead is roughly $O(n)$. For $\delta_{\rm SP}^2 n\gg 1$, the additional overhead is $O(\delta_{\rm SP}^2 n^2)$.

We remark that if one wants to further perform mitigation on gate noise, one can do that on top of the mitigated distribution $\hat{\bm{P}}$.

\subsection{Results from simulator}

\begin{figure}[h]
    \centering
\[\Qcircuit @C=.7em @R=.7em @! {
\lstick{\ket{0}_\text{1}}&\gate{R_y(\theta)}& \ctrl{1}\qw    & \meter  \\
\lstick{\ket{0}_\text{2}}&\gate{R_y(\theta)}& \targ       & \meter  
}\]
    \caption{The 2-qubit benchmark circuit for comparing two mitigation schemes CM and SM.}
    \label{fig:circ_sim2q}
\end{figure}

\begin{figure}[h]
    \centering
\[\Qcircuit @C=.7em @R=.7em @! {
\lstick{\ket{0}_\text{1}}&\gate{R_y(\theta)}&\qw        & \targ    & \meter  \\
\lstick{\ket{0}_\text{2}}&\gate{R_y(\theta)}&\ctrl{1}\qw& \ctrl{-1}\qw       & \meter\\  
\lstick{\ket{0}_\text{3}}&\gate{R_y(\theta)}&\targ      & \ctrl{1}\qw    & \meter  \\
\lstick{\ket{0}_\text{4}}&\gate{R_y(\theta)}&\qw        & \targ       & \meter\\ 
}\]
    \caption{The 4-qubit benchmark circuit for comparing two mitigation schemes CM and SM.}
    \label{fig:circ_sim4q}
\end{figure}

Here, we compare mitigating the state preparation error and measurement error separately and together for a simple circuit shown in Fig.~\ref{fig:circ_sim2q}. (A four-qubit example circuit that will be tested later is shown in Fig.~\ref{fig:circ_sim4q}.) The $R_y$ gate is defined as $R_y(\theta)=\exp(-i \theta Y/2 )$ and the ideal output state is 
\begin{eqnarray}\nonumber
    \ket{\hat{\psi}}&=&\cos^2 \frac{\theta}{2} \ket{00}+ \sin^2 \frac{\theta}{2}\ket{01}\\
    &~&+\sin \frac{\theta}{2}\cos \frac{\theta}{2}\ket{10}+\sin \frac{\theta}{2}\cos \frac{\theta}{2}\ket{11}.
\end{eqnarray}

Since the measurement can only give probability distributions instead of wave functions, we define the overlap (or fidelity) between two probability distributions as~\cite{nielsen2010quantum}  
\begin{equation}
    F(\bm{P},\bm{P}')=\left(\sum_{i=1}^{2^n} \sqrt{p_ip_i'}\right)^2\,,
\end{equation}
where $p_i$ is the probability in the distribution $\bm{P}$ for the state $\ket{i}$.  The above definition is related to their Heilinger distance~\cite{hellinger1909neue} $H(\bm{P},\bm{P'})$ via  $\sqrt{F(\bm{P},\bm{P'})}=1-H^2(\bm{P},\bm{P'})$ between the two distributions. The fidelity of an output distribution $\bm{P}_{\theta}$ is then defined as the overlap with the ideal probability distribution $\hat{\bm{P}}$
\begin{equation}
    f(\theta)=F(\bm{P}_{\theta},\hat{\bm{P}}),
\end{equation}
where $\hat{\bm{P}}=(\cos^4\frac{\theta}{2},\sin^4\frac{\theta}{2},\frac{1}{4}\sin^2\theta,\frac{1}{4}\sin^2\theta)$.  We use a simulator with a state preparation error $\delta_{\rm SP}=0.05$ and $\delta_M^0=0.04, \delta_M^1=0.06$ for both qubits.

The theoretical calculation is straightforward but gives rise to messy expressions for both mitigation schemes, except for $\theta=0$, at which we have
\begin{gather}
    f_{\rm CM}(0)=1-\frac{\delta_{\rm SP}^{(1)}}{2},\\
    f_{\rm SM}(0)=1-\frac{4}{3}\delta_{\rm SP}^{(1)}\delta_{\rm SP}^{(2)},
\end{gather}
and, at $\theta=\pi /2$, we have
\begin{gather}
    f_{\rm CM}(\frac{\pi}{2})=f_{\rm SM}(\frac{\pi}{2})=1.
\end{gather}
Therefore, we expect that, near $\theta=0$, the two mitigation schemes achieve fidelity with the ideal distribution at different orders, with the CM infidelity roughly approaching the first order $O(0.05)$ and SM infidelity approaching the second order $O(0.05^2)$ in most region. The simulation results for both the two-qubit and four-qubit circuits are shown in Fig.~\ref{fig:dspm_sim}, where SM has better fidelity and less fluctuation in a large proportion of the parameter region in both cases.

\subsection{Experimental demonstration on superconducting quantum computers}
\label{sec:experimental}

\begin{figure}[h]
    \centering
    \includegraphics[width=0.99\linewidth]{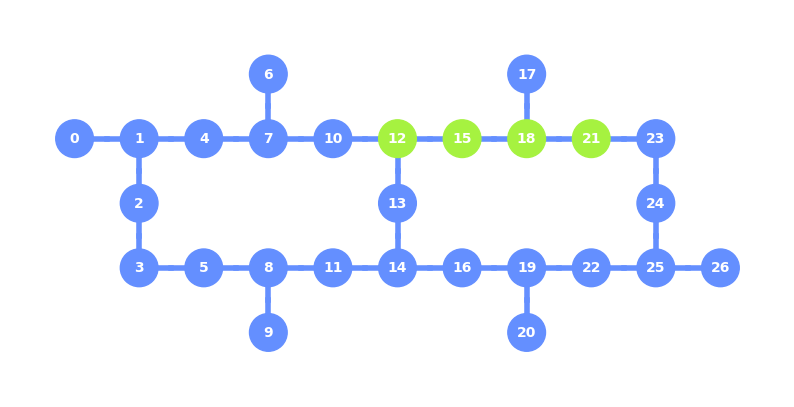}
    \caption{The qubit geometry for ibmq\_mumbai. We run the benchmark circuit on qubits [21,18,15,12] (highlighted in green). }
    \label{fig:mumbai}
\end{figure}

In this section, we experimentally implement the $\delta_{\rm SP}$ characterizing circuits and obtain the estimated values of $\delta_{\rm SP}$ on superconducting quantum computers of IBM's cloud platform. Then, we perform quantum state tomography (with readout mitigation) to confirm that the $\delta_{\rm SP}$ values we obtain match the tomography results. Finally, we compare the two mitigation schemes on a 4-qubit circuit and demonstrate that the SM obtains better probability fidelity than the CM.


To demonstrate the direct application to circuit runs on current NISQ devices, we use the circuit shown in Fig.~\ref{fig:circ_sim4q} to benchmark both mitigation schemes. We choose $\theta=\pi/5$. We run the circuit on the ibmq\_mumbai with the initial qubit layout [21,18,15,12]. Using the technique shown in Sec.~\ref{sec:dspnoise}, we obtained the state preparation errors $\{\delta_{\rm SP}\}=\{0.007(3),0.007(2),0.007(1),0.029(1)\}$. Then, we carry out the procedure for the CM and SM, respectively. The raw and mitigated probability results are shown in Fig.~\ref{fig:exprmt4q}. The CM gives 99.6(4)\% probability fidelity while the SM gives 99.9(2)\% probability fidelity. The simulation without gate noise, but with state-preparation errors and measurement errors, the same as the experimental data, indicates a 99.6273\% to 99.9998\% improvement in fidelity, close to the experimental results.

\section{Conclusion}
\label{sec;conclusion}

We have proposed a resource-efficient approach to separately quantify the state preparation and readout error rates. With these two errors separately quantified, we have also described efficient methods to mitigate them, especially mitigating state preparation errors with linear complexity.  In addition to theoretical analysis, we have additionally presented results from cloud experiments on IBM's superconducting quantum computers, and the results well match our theoretical predictions.  

In this work, we have assumed high fidelity of X gates, which is usually the case in real quantum devices~\cite{kim2023evidence}. We also proposed an efficient way to obtain $\delta_{\rm SP}$ in the presence of CNOT gate noise, which, in the experiment, gives consistent results compared to those obtained by the ZNE and PEC. Moreover, we believe our methods can be easily extended beyond superconducting quantum computers to other physical systems.

The experimental results show practical applications of our quantification and mitigation scheme for $\delta_{\rm SP}$ on current noisy quantum computers. Although the improvement of fidelity is limited due to the relatively small $\delta_{\rm SP}$ compared to the $\delta_{M}$ on IBM's superconducting quantum computers, the improvement would be much evident when the $\delta_{M}$ is reduced to a similar scale as $\delta_{\rm SP}$ in the future. We advocate that in reporting the properties of quantum bits, the state preparation error rate should also be provided as an important metric (at least the portion in the incoherent mixture of $|1\rangle\langle1|$ when preparing $|0\rangle$; but, in principle, quantum state tomography can be used for further characterization).

The work focuses on single-qubit errors, which can be extended to multi-qubit correlated errors. One will need to add twirling to the initial state preparation so that its error becomes incoherent, which will require an 
 exponentially increasing number of circuits. One can also straightforwardly replace all the error assignment matrices with the full $2^n \times 2^n$ matrix, although this will not be practical for large $n$. However, one may consider dividing the system into groups of qubits that have significant correlations and then characterizing different groups independently, similar to the two-qubit case in~\cite{bravyi2021mitigating}. One may also implement the truncation schemes in~\cite{nation2021scalable} to reduce the complexity of mitigation further. In general, methods to efficiently deal with multiqubit-correlated errors in state preparation are worth exploring in the future.

\smallskip \noindent {\bf Code availability}. The codes and data for the cloud experiments in this work are available upon reasonable request. The codes for demonstrating separate quantification are available at \href{https://github.com/HongyeY/StatePrepError_pub}{github}.


\acknowledgments We thank Yusheng Zhao and Nicholas Bronn for many useful discussions.
This work was partly supported by the National Science Foundation under Grant No. PHY 2310614, in particular, on the part of the state preparation error's characterization and mitigation, and by the U.S. Department of Energy, Office
of Science, National Quantum Information Science Research
Centers, Co-design Center for Quantum Advantage (C2QA)
under Contract No. DE-SC0012704, in
particular, on the part of the algorithmic cooling.
This research also used resources from the Oak Ridge Leadership
Computing Facility, which is a DOE Office of Science User
Facility supported under Contract DE-AC05-00OR22725, and
the Brookhaven National Laboratory operated IBM-Q Hub.
The results presented in this work do not reflect the views of
IBM and its employees.

\bibliographystyle{quantum}
\bibliography{MitiSP}

\clearpage

\appendix
\section{Comparison and Extension of the Laflamme-Lin-Mor algorithmic-cooling scheme for separate quantification of state preparation and readout errors}
\label{app:LLM}

In this appendix, we briefly review the algorithmic-cooling  method proposed in \cite{laflamme2022algorithmic} and compare it with our $\delta_{\rm SP}$-characterization scheme.

\subsection{Brief summary of algorithmic-cooling scheme for quantifying $\delta_{\rm SP}$}

The measurement-based algorithmic-cooling method \cite{laflamme2022algorithmic} is based on algorithmically discarding the higher-energy state $\ket{1}$ from the measurement outcome of ancillary qubits, so that it can cool the system closer to the ground state $\ket{0}$ (see Fig.~\ref{fig:circ_ac}).

\begin{figure}[h]
    \centering
    
    \[\Qcircuit @C=.7em @R=.2em @! {
\lstick{\ket{0}_\text{a}}& \targ     & \meter  &\control \cw\\
\lstick{\ket{0}_\text{t}}& \ctrl{-1}\qw  & \qw  & \gate{\text{discard}} \cwx
}\]

    \caption{Circuits for algorithmic-cooling scheme with one ancilla, where the results of target qubit is discarded if the ancilla qubit is measured at $\ket{1}$.}
    \label{fig:circ_ac}
\end{figure}

We assume $\delta_{\rm SP}=\delta$ for all qubits in this section for simplicity. One can verify that the effective state-preparation error is reduced to $O(\delta^2)$ after discarding all the results where the ancilla qubit is measured at $\ket{1}$. Such procedure can be further extended to $k$ ancilla qubits, where each ancilla qubit is prepared at $\ket{0}$ (with error $\delta$), connected with the target qubit with a CNOT gate similar to Fig.~\ref{fig:circ_ac}. After discarding all the results where any ancilla qubit is measured at $\ket{1}$, the effective state-preparation error for $q_t$ is reduced to $O(\delta^k)$. If $k$ is sufficiently large, the effective SPAM error of $q_t$ is almost purely measurement errors, thus can be measured directly. Then one can solve $\delta_{\rm SP}$ according to Eq.~\eqref{eq:SPAM=SP+M}, where $\delta_{\rm SPAM}$ can be measured directly and $\delta_{M}$ can be measured after discarding $\ket{1}_a$ results from large-$k$ algorithmic-cooling circuit.

Compared to our $\delta_{\rm SP}$-characterization scheme, the algorithmic-cooling scheme focuses more on reducing the effects of $\delta_{\rm SP}$ via post-selection from measuring ancilla qubits. After $\delta_{\rm SP}$ is reduced to almost 0, the measurement error can be directly measured, and $\delta_{\rm SP}$ can then be solved. In our characterization scheme, the state-preparation error and measurement error are both obtained by directly solving the linear equations Eqs.~\eqref{eq:spamT} and~\eqref{eq:SPAM=SP+M}. However, the algorithmic-cooling scheme requires multiple ancilla qubits, with multiple CNOT gates connected to the target qubit, and, in practice, contains more noise and errors from noisy CNOT gates and is hard to implement if the target qubit is not physically connected with multiple qubits. In our $\delta_{\rm SP}$-characterization scheme, we only require one additional ancilla qubit and one CNOT gate so that the noise from CNOT is much easier to deal with by gate-noise mitigation schemes. Thus, our scheme is more practical to be implemented on current noisy quantum computers.

\subsection{Quantifying $\delta_{\rm SP}$ via algorithmic-cooling with one ancilla}
In this section, we show that it is possible to extract the state preparation error rate in the algorithmic-cooling scheme with one ancilla. We assume for simplicity that the readout error is symmetric (i.e., $\delta_{\rm M}^0=\delta_{\rm M}^1$) as in Ref.~\cite{laflamme2022algorithmic} or else we can use the symmetrized readout protocol. We will use scripts to label the two qubits 1 and 2. First, we use 1 as the control and 2 as the target and go through the procedure of post-selecting `0' for qubit 2. This reduces the state preparation error of qubit 1 to $\tilde{\delta}_{\rm SP,1}=\delta_{\rm SP,1} f_{12}$,
 where  the factor $f_{12}$ was obtained and defined in  Ref.~\cite{laflamme2022algorithmic}, and we quote it here,
 \begin{equation}
 f_{12}\equiv\frac{2(\delta_{\rm SP,1}+\delta_{\rm M,1}-2\delta_{\rm SP,1}\delta_{\rm M,1})}{1+(1-2\delta_{\rm SP,1})(1-2\delta_{\rm M,1})(1-2\delta_{\rm SP,2})},
 \end{equation}
 where the subscripts are used to denote which qubits.
 The total measured SPAM error (after post-selecting qubit 2) is $\tilde{\delta}_{\rm SPAM,1}=\tilde{\delta}_{\rm SP,1}+\delta_{\rm M,1}$.

 Next, we use 2 as the control and 1 as the target and go through the procedure of post-selecting `0' for qubit 1. This reduces the state preparation error of qubit 2 to $\tilde{\delta}_{\rm SP,2}=\delta_{\rm SP,2} f_{21}$,
 where
 \begin{equation}
 f_{21}\equiv\frac{2(\delta_{\rm SP,2}+\delta_{\rm M,2}-2\delta_{\rm SP,2}\delta_{M,2})}{1+(1-2\delta_{\rm SP,2})(1-2\delta_{\rm M,2})(1-2\delta_{\rm SP,1})}.
 \end{equation}
 The total measured SPAM error (after post-selecting qubit 2) is $\tilde{\delta}_{\rm SPAM,2}=\tilde{\delta}_{\rm SP,2}+\delta_{\rm M,2}$.
 Together with two other measurable quantities,
 $\delta_{\rm SPAM,i}=\delta_{\rm SP,i}+\delta_{\rm M,i}-2\delta_{\rm SP,i}\delta_{M,i}$ (for $i=1,2$), we have a total of four equations for four unknowns, with which one can solve in principle.
 
 We have performed numerical tests and confirmed that the above equations can be solved.
For example, if we measure the four SPAM errors (using the symmetrized readout protocol to average the two readout errors) and obtain the measured results:
 \begin{eqnarray} 
 &&\{\delta_{\rm SPAM,1},\delta_{\rm SPAM,2},\tilde{\delta}_{\rm SPAM,1},\tilde{\delta}_{\rm SPAM,2}\}\nonumber\\
 &&=\{0.22923, 0.277084, 0.222717, 0.275828\}.
 \end{eqnarray}
 Then, solving the above equations leads to a single solution: 
 \begin{eqnarray}
 \!\!\!\!    && \{\delta_{\rm SP,1},\delta_{\rm M,1},{\delta}_{\rm SP,2},{\delta}_{\rm M,2}\}\nonumber\\
  \!\!\!\!   &&=\{0.0242911, 0.215404, 0.0172848, 0.269102\}.
      \end{eqnarray}
      With additional characterization of the asymmetry in the readout, we can obtain a complete and separate characterization of SPAM errors.

      We note that the algorithmic cooling approach here requires solving coupled nonlinear equations, whereas the method presented in the main text requires solving only linear equations.

\section{Vector representations of single qubit spin-flip error and the proof for the $\delta_{\rm SP}$ equation \eqref{eq:spamT}}\label{app:vecErr}
Consider a bit-flip error, with $\delta^0$ probability flipping the state $\ket{0}$ to $\ket{1}$ and $\delta^1$ for $\ket{1}$ to $\ket{0}$. We can define a vector for such error $D\equiv\{\delta^0,\delta^1\}$, with a map $f(D)=A$ to its corresponding assignment matrix 
\begin{eqnarray}
    f(D)=f(\{\delta^0,\delta^1\})=A=\begin{pmatrix}
        1-\delta^0 & \delta^1\\
        \delta^0 & 1-\delta^1
    \end{pmatrix}.
\end{eqnarray}
The combination of error $D_1$ on top of error $D_2$ is defined as the matrix product of their corresponding assignment matrices $A_1$ and $A_2$,
\begin{equation}
    D_1 \circ D_2\equiv f^{-1}(f(D_1)f(D_2))=f^{-1}(A_1 A_2)
\end{equation}
or written explicitly
\begin{eqnarray}\nonumber
    D_1 \circ D_2&=&\{\delta^0_1,\delta^1_1\}\circ \{\delta^0_2,\delta^1_2\}\\\nonumber
    &\equiv& \left\{\delta^0_1+\delta^0_2(1-\delta^0_1-\delta^1_1),\right.\\
    &&\left.\delta^1_1+\delta^1_2(1-\delta^0_1-\delta^1_1) \right\}.
\end{eqnarray}
Note that $A$ is invertible \textit{iff} $\delta^0+\delta^1\neq 1$. Thus, error vectors with the combination rule $(\{D, \delta^0+\delta^1\neq 1\},\circ)$ form a group. $D_a \circ D_b$ is usually not equal to $D_b \circ D_a$, unless $\delta^0_a/\delta^1_a=\delta^0_b/\delta^1_b$. Particularly, for those errors satisfy $\delta^0=\delta^1$ (e.g. state-preparation errors  \eqref{eq:ASP}), they commute with each other. We can derive the equations \eqref{eq:spamT} as follows
\begin{eqnarray}\nonumber
    \tilde{D}_{\rm SPAM}^a&=&D_M^a\circ D_{\rm SP}^t \circ D_{\rm SP}^a\\\nonumber
    &=&D_M^a \circ D_{\rm SP}^a\circ D_{\rm SP}^t\\
    &=&D_{\rm SPAM}^a\circ D_{\rm SP}^t,
\end{eqnarray}
where we use the relation $D_{\rm SPAM}^a=D_M^a \circ D_{\rm SP}^a $. By measuring $D_{\rm SPAM}^a$ and $\tilde{D}_{\rm SPAM}^a$, one can solve the $D_{\rm SP}^t=(D_{\rm SPAM}^a)^{-1} \circ \tilde{D}_{\rm SPAM}^a$.

\section{Obtain $\delta_{\rm SP}$ for general uncorrelated CNOT gate noise}\label{app:moreCNOT}
For general CNOT gate noise, we consider the case without correlating error, where we assume the noise is effectively a channel applied after the CNOT operation. For general noise channels, one can apply Pauli-twirling to effectively turn the noise channels into Pauli channels. Then the single-qubit error channel can be written as a direct product of two Pauli-like single-qubit channels
\begin{equation}
\Lambda(\rho)=\sum_{a \in \{I,X,Y,Z\}} p_a P_a(\rho) P_a.
\end{equation}
With same argument, by adding a random Pauli Z operator with 50\% probability for each qubit before the CNOT gate, we ensure that the state-preparation error produces effectively $\rho=  (1-\delta_{\rm SP} )|0\rangle\langle0|+ \delta_{\rm SP}|1\rangle\langle1|$, on which the Pauli Z channel has no effect but other Pauli channels act effectively like the bit-flip error. For a CNOT gate with qubit $q_2$ controlling $q_1$, the error assignment matrix of $q_1$ can be written as
\begin{equation}\label{eq:ACNOT}
    A_{\rm CNOT}^{(q_2,q_1)q_1}=\begin{pmatrix}
        1-\delta_{\rm CNOT}^{(q_2,q_1)q_1} & \delta_{\rm CNOT}^{(q_2,q_1)q_1}\\
        \delta_{\rm CNOT}^{(q_2,q_1)q_1} & 1-\delta_{\rm CNOT}^{(q_2,q_1)q_1}
    \end{pmatrix}.
\end{equation}
With this additional error, the procedure of quantifying $\delta_{\rm SP}$ becomes more complicated. The circuits we use are as follows:
\begin{figure}[h]
    \centering
    
    \[\Qcircuit @C=.7em @R=.7em @! {
\lstick{\ket{0}_{q_1}}& \targ     & \meter  & \rstick{\tilde{D}_{\rm SPAM,q_1}^{(q_2,q_1)q_1}} \qw\\
\lstick{\ket{0/1}_{q_2}}& \ctrl{-1}\qw  & \meter  & \rstick{D_{\rm SPAM,q_2}^{(q_2,q_1)q_2}} \qw
}\]

\[\Qcircuit @C=.7em @R=.7em @! {
\lstick{\ket{0/1}_{q_1}}& \targ     & \targ     & \meter  & \rstick{\tilde{D}_{\rm SPAM,q_1}^{(q_2,q_1)_2 q_1}} \qw\\
\lstick{\ket{0}_{q_2}}& \ctrl{-1}\qw& \ctrl{-1}\qw  & \meter  &  \qw
}\]

    \caption{Circuits for measuring $\delta_{\rm SP}$ with CNOT gate noise.}
    \label{fig:circ_dspCNOT}
\end{figure}

We can get the following relations from the circuits
\begin{gather}
\label{eq:cnot1}
    \tilde{D}_{\rm SPAM,q_1}^{(q_2,q_1)q_1}=D_{\rm SPAM}^{q_1} \circ D_{\rm CNOT}^{(q_2,q_1)q_1} \circ D_{\rm SP}^{q_2}\\\label{eq:cnot2}
    D_{\rm SPAM,q_2}^{(q_2,q_1)q_2}=D_{\rm SPAM}^{q_2} \circ D_{\rm CNOT}^{(q_2,q_1)q_2} \\\label{eq:cnot3}
    \tilde{D}_{\rm SPAM,q_1}^{(q_2,q_1)_2q_1}=D_{\rm SPAM}^{q_1} \circ (D_{\rm CNOT}^{(q_2,q_1)q_1})^2 \circ D_{\rm CNOT}^{(q_2,q_1)q_2}
\end{gather}
where $D_{\rm SPAM}^{q_1}$ and $D_{\rm SPAM}^{q_2}$ can be directly measured. From the equation \eqref{eq:cnot2} we can solve the $D_{\rm CNOT}^{(q_2,q_1)q_2}$. Then by inserting it into \eqref{eq:cnot3} we can get $D_{\rm CNOT}^{(q_2,q_1)q_1}$. Finally, we insert $D_{\rm CNOT}^{(q_2,q_1)q_1}$ into \eqref{eq:cnot1} and get the $D_{\rm SP}^{q_2}$.

\begin{figure*}[h]
\begin{tabular}{ll}
  (a)  & (b)  \\ 
  \includegraphics[width=0.48\textwidth]{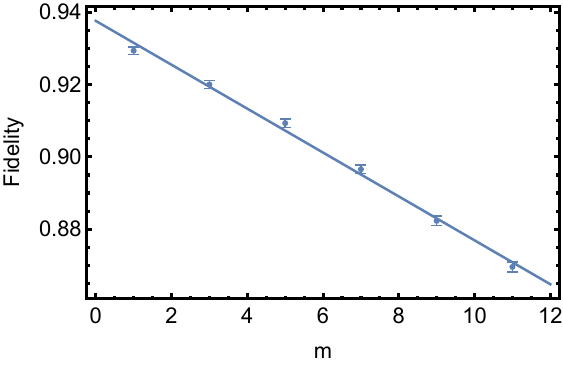} &  \includegraphics[width=0.48\textwidth]{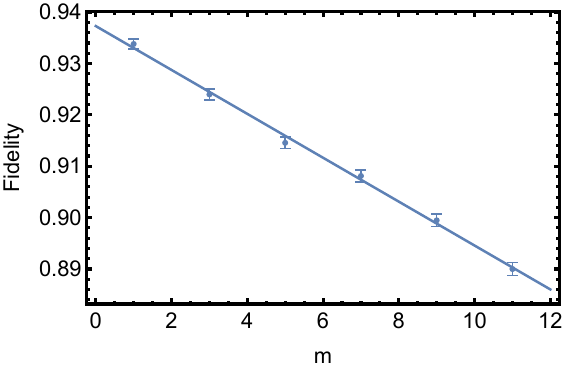}\\
  (c) & (d)\\
\includegraphics[width=0.48\textwidth]{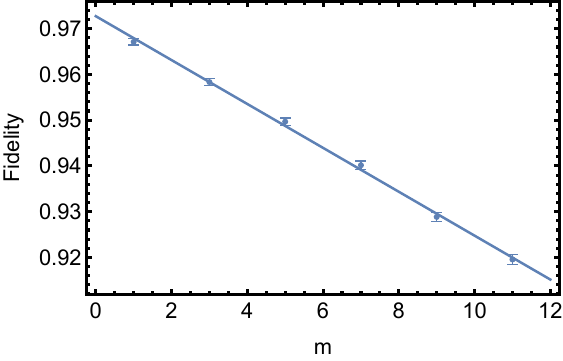} 
&
\includegraphics[width=0.48\textwidth]{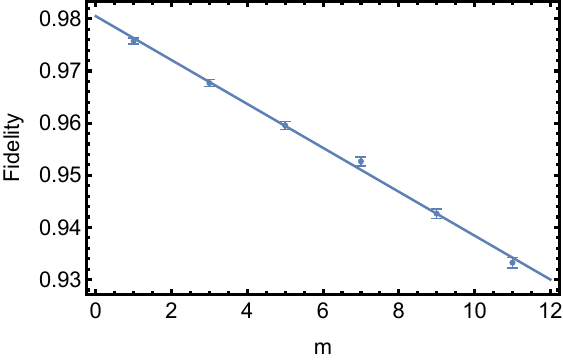}
\end{tabular}
\caption{\label{fig:ZNE}  Zero-noise extrapolation for $\delta_{\rm SP}$ characterization circuit with $q_{12}, q_{10}$ on ibmq\_mumbai. Each circuit is repeated 64000 times. (a) Fidelity of $q_{10}$ with initial state $\ket{00}$ for circuit \ref{fig:circ_dsp} with $q_{12}$ as target qubit. The fitted result gives $1-\tilde{\delta}_{\rm SPAM}^{10,0,\rm miti}=0.938(1)$. (b) Fidelity of $q_{10}$ with initial state $\ket{01}$ for circuit \ref{fig:circ_dsp} with $q_{12}$ as target qubit. The fitted result gives $1-\tilde{\delta}_{\rm SPAM}^{10,1,\rm miti}=0.937(1)$. (c) Fidelity of $q_{10}$ with initial state $\ket{00}$ for circuit \ref{fig:circ_dsp} with $q_{10}$ as target qubit. The fitted result gives $1-\delta_{\rm SPAM}^{10,0,\rm miti}=0.973(1)$. (d) Fidelity of $q_{10}$ with initial state $\ket{01}$ for circuit \ref{fig:circ_dsp} with $q_{10}$ as target qubit. The fitted result gives $1-\delta_{\rm SPAM}^{10,0,\rm miti}=0.980(1)$.
  }
\end{figure*}

\section{Detailed comparison for calculating $\delta_{\rm SP}$ with and without CNOT gate noise extrapolation}\label{app:CNOT}

In this appendix, we compare the estimation of $\delta_{\rm SP}$ from noisy-CNOT results~\eqref{eq:dspnoisy} and mitigated-CNOT results. To mitigate CNOT gate noise, we use the zero-noise extrapolation technique~\cite{kandala2019error}, where CNOT gates are repeated $m=1,3,5,...$ times to get different noise-scale results. After that, we use linear extrapolation to extrapolate them to zero-noise ($m=0$) point and get the mitigated data. The extrapolating results for calculating $\delta_{\rm SP}^{(12)}$ via ancilla qubit $q_{10}$ of ibmq\_mumbai are shown in Fig.~\ref{fig:ZNE}. From the mitigated results, we get the $\delta_{\rm SP}^{(12)}=0.041(4)$, which is very close to the results from unmitigated data $\delta_{\rm SP}^{(12)}=0.042(2)$ (shown in Sec.~\ref{sec:dspnoise}).

\end{document}